\documentclass[mathleft,fleqn,%
]{an}
\usepackage{graphicx}
\usepackage[varg]{txfonts}
\newcommand{\lsun}{${L}_{\odot}$}
\newcommand{\msun}{${M}_{\odot}$}
\newcommand{\rsun}{${R}_{\odot}$}
\newcommand{\kms}  {km~s$^{-1}$}

\begin{document}

\Pagespan{1}{}
\Yearpublication{2020}%
\Yearsubmission{2019}%
\Month{0}%
\Volume{999}%
\Issue{0}%
\DOI{asna.202000000}%

\title{Flickering of the jet-ejecting symbiotic star MWC 560}

\author{R. K. Zamanov\inst{1}\fnmsep\thanks{Corresponding author:       {rkz@astro.bas.bg}}
\and  S. Boeva\inst{1}
\and  K. A. Stoyanov$^{1}$
\and  G. Latev$^{1}$
\and  B. Spassov$^{1}$ 
\and  A. Kurtenkov$^{1}$
\and  G. Nikolov$^{1}$
}
\titlerunning{Flickering of MWC 560}
\authorrunning{Zamanov, Boeva, Stoyanov et al.}
\institute{
Institute of Astronomy and National Astronomical Observatory, Bulgarian Academy of Sciences, Tsarigradsko Shose 72, \\
BG-1784 Sofia, Bulgaria
}

\received{2019 December 3}
\accepted{2020 March 31}


\abstract{%
We analyse optical photometric data of short term variability (flickering) 
of the accreting white dwarf in the jet-ejecting symbiotic star  MWC~560.
The observations are obtained in  17 nights during the period  November 2011 - October 2019.  
The colour-magnitude diagram shows 
that the hot component of the system becomes redder as it gets brighter.
For the flickering source we find that it has colour $0.14 < B-V < 0.40$,    
temperature in the range  $6300 < T_{fl} < 11000$~K, 
and radius  $1.2 < R_{fl}  < 18$~\rsun. 
We find  a strong correlation 
(correlation coefficient 0.76, significance $< 0.001$) between B band magnitude
and the average radius of the flickering source --
as the brightness of the system increases  the size of the flickering source also increases. 
The estimated temperature is similar to that of the bright spot of cataclysmic variables.
In 2019 the flickering is missing, and the B-V colour of
the hot component becomes bluer. }
\maketitle

\section{Introduction}
The symbiotic stars  are wide binaries with long orbital periods 
(from 100 days to 100 years) in 
which material is transferred from an evolved red giant star to an white dwarf
or a neutron star (Miko{\l}ajewska 2012).
The symbiotic star MWC~560 (V694 Mon) was identified
as an emission line object at the Mount Wilson observatory spectroscopic surveys   
(Merrill  \&  Burwell 1943). 
The spectroscopic observations of MWC~560 in 1984 showed  that its an extraordinary 
symbiotic star with absorption extending out to 
$-3000$~\kms \ at H$\beta$ and higher members of the Balmer series (Bond et al. 1984). 
In early 1990 the outflow velocities  reached
6000 -- 7000~km~s$^{-1}$ (Tomov et al. 1990; Szkody, Mateo \& Schmeer 1990).
Tomov et al. (1990) proposed that the observed absorptions are caused by 
a collimated outflow  along the line of sight --
a low-energy  analog of the jets of the microquasar SS~433.
The outflow may be a highly-collimated baryon-loaded jet (Schmid et al. 2001) 
or a wind from the polar regions (Lucy, Knigge \& Sokoloski 2018). 
MWC~560 is considered to be a non-relativistic analog of the quasars not only because of its jets, 
but also to the resemblance of its emission
lines to that of the low-redshift quasars (Zamanov \& Marziani 2002) 
and the absorption lines  to that of the broad absorption lines quasars (Lucy et al. 2018). 
The orbital period of the binary is thought to be $P_{orb} = 1931 \pm 162$~d (Gromadzki et al. 2007) although 
recently Munari et al. (2016) supposed that it could be considerably shorter, $P_{orb} \approx 330.8$~d.

Systematic searches  for  flickering variability in  symbiotic stars 
and related objects (Dobrzycka, Kenyon \& Milone 1996;
Sokoloski, Bildsten \& Ho \ 2001;  Gromadzki et al. 2006; 
Angeloni et al. 2013)  have shown that 
optical flickering  is a  rarely detectable  phenomenon in symbiotic stars. 
Among more than 200 symbiotic stars known, 
only in 11 objects flickering activity is  visible.
A flickering  variability of MWC~560 
of up to 0.2 mag on timescale of a few minutes  was first reported by Bond et al. (1984). 
The amplitude is in the range 0.1 -- 0.7 mag and the detected 
quasi-periods are from 11 to 160 min (Tomov et al. 1996). 
The intra-night variability was a persistent feature till 2018, when  the variability on time-scale of minutes 
became undetectable (Goranskij et al. 2018).

Here, we report quasi-simultaneous observations of the flickering variability of 
the jet-ejecting symbiotic star MWC~560 (most of them in the two optical bands -- B and V)
and
analyze the colour changes, temperature and radius of the flickering source
and their response to the brightness variations.

\section{Observations and data analysis}

The observations were performed with four telescopes equipped with CCD cameras: 
\begin{itemize}
  \item the 2.0 m telescope of the National Astronomical Observatory (NAO) Rozhen, Bulgaria  
  (Bonev \& Dimitrov 2010)  
  \item the 50/70 cm Schmidt telescope of NAO Rozhen    
  \item the 60 cm telescope of NAO Rozhen  
  \item the 60 cm telescope of the Belogradchik Observatory, Bulgaria (Strigachev \& Bachev 2011)  
\end{itemize}
The data reduction was done with IRAF (Tody 1993) following standard recipes for processing of CCD images and aperture photometry. 
A few comparison stars from the list of Henden \& Munari (2006) and APASS DR9 have been used. 
The typical photometric errors are 0.007~mag in U-band, 
0.005~mag  in B-band, and 0.004~mag in V-band.

\begin{figure*}    
   \vspace{7.0cm}     
   \includegraphics{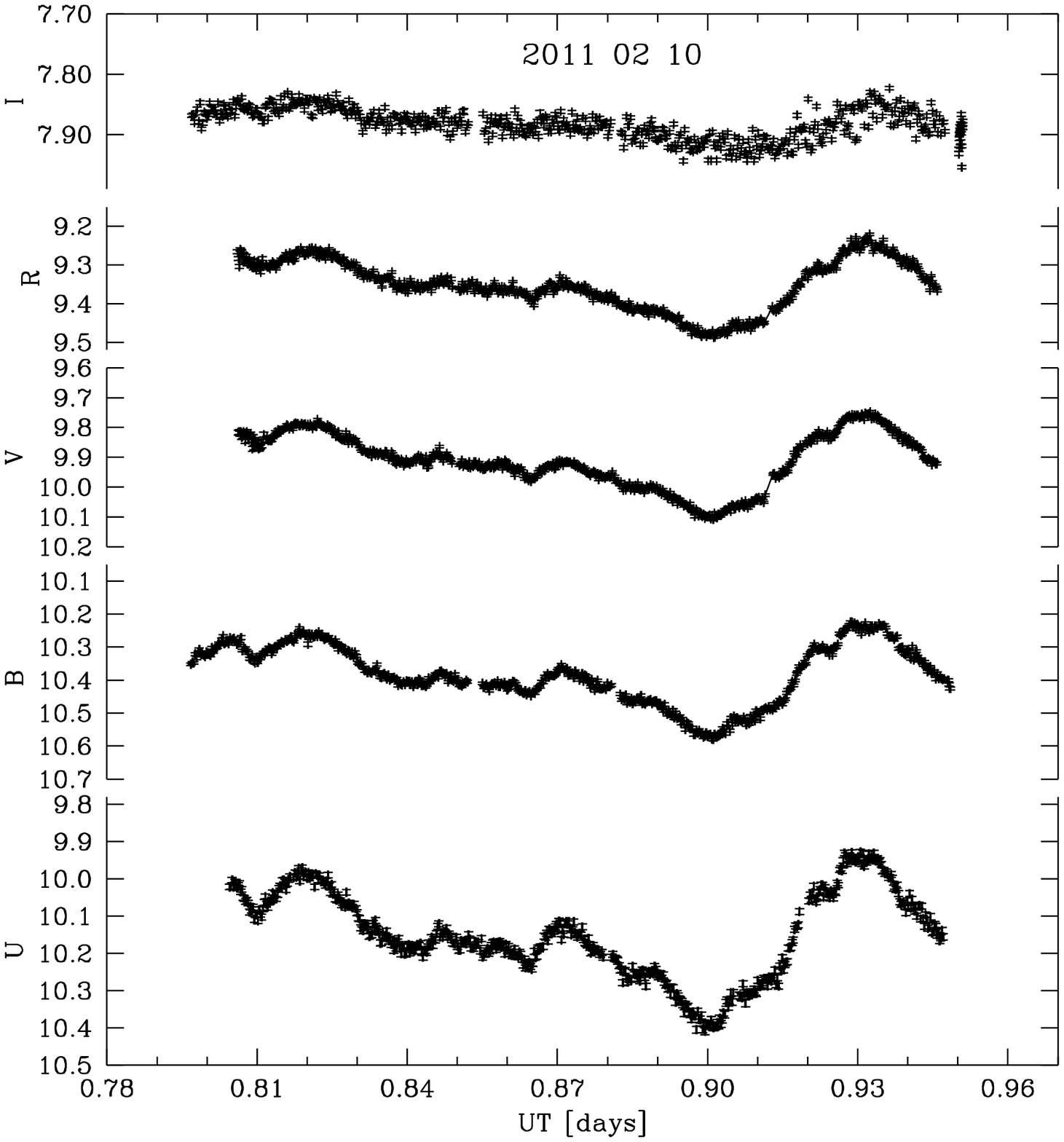}      
   \includegraphics{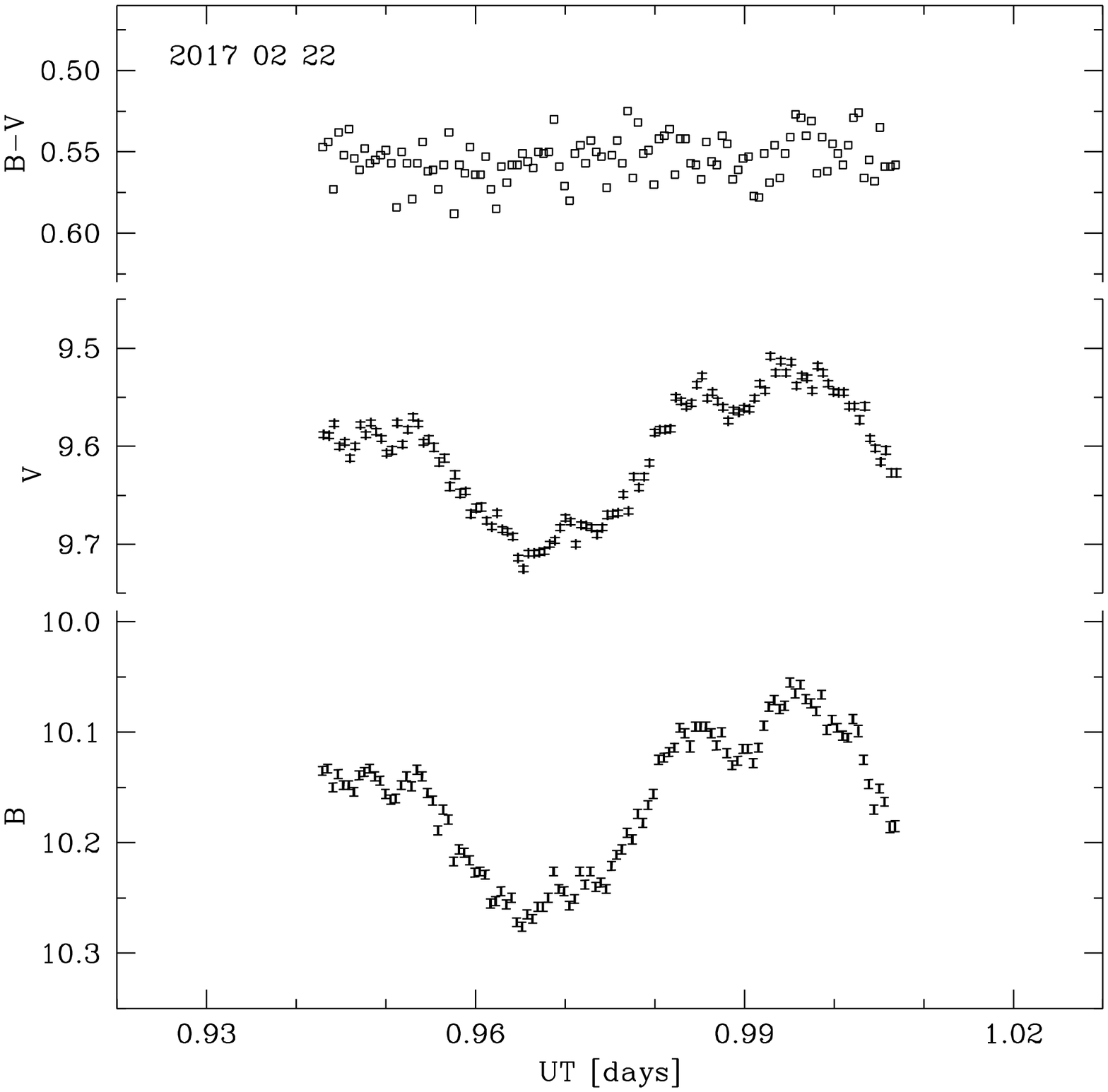}  
   \includegraphics{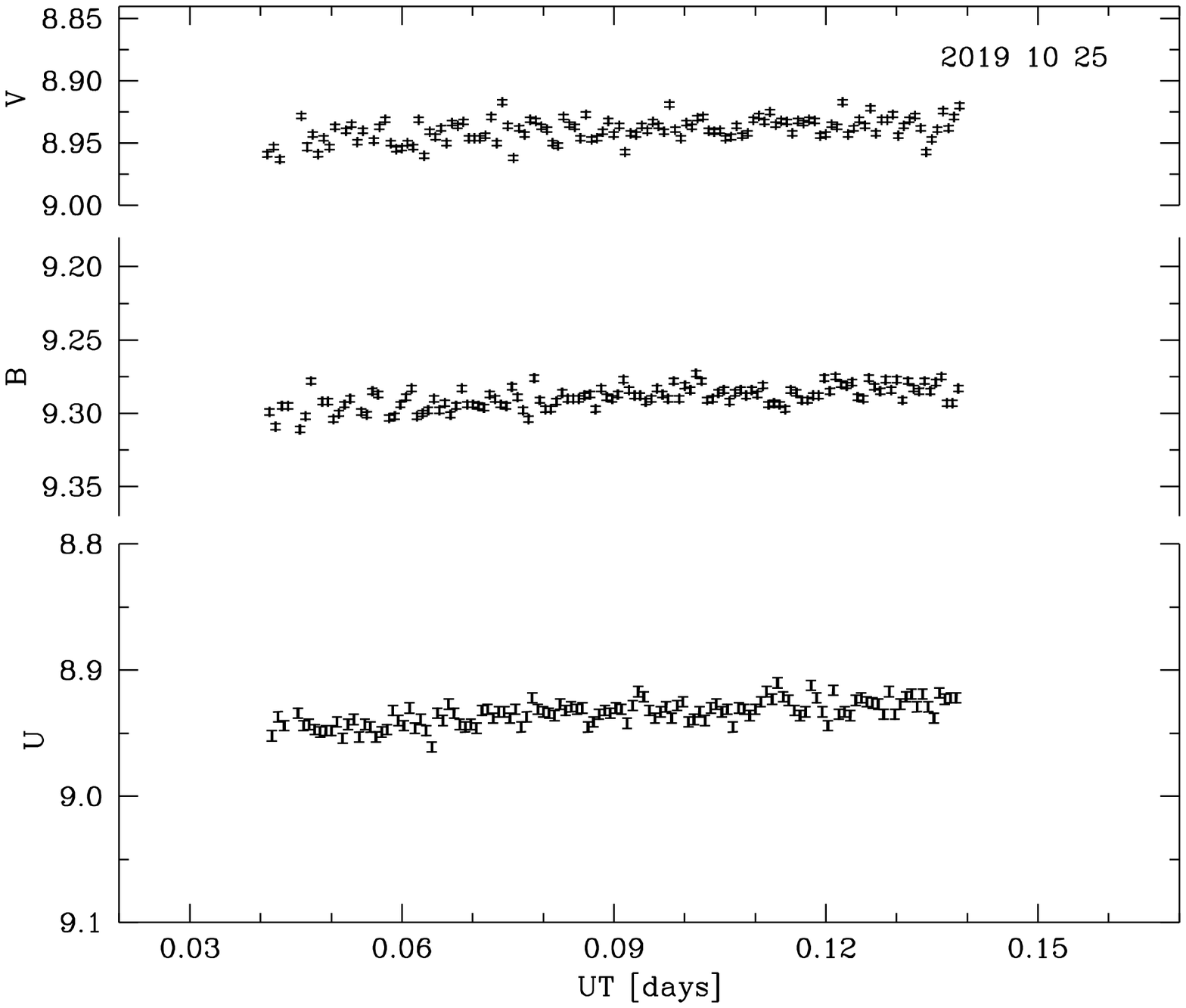}  
   \caption[]{Examples of the short term variability of  MWC~560. The label in each panel indicate the date of
   observations in format YYYYMMDD. 
   The left panel is 20110210 --  UBVRI data (the flickering is visible even in I band, 
   its amplitude is decreasing to the red bands).
   The middle panel is  20170222 -- the light curves in B and V bands 
   together with the calculated B-V colour.
   The right panel is 20191025 - UBV data,  the flickering is missing. }
   \label{fig.ex}      
 \end{figure*}

Three examples of our data are given on Fig.~\ref{fig.ex}. 
On the left panel (20110210) are plotted  UBVRI data.
The amplitude in U-band is 0.47 mag, in B-band - 0.34 mag, in V-band - 0.32 mag, in R-band - 0.23 mag, in I-band -- 0.07 mag.  
The amplitude of the flickering is decreasing to longer wavelengths, mainly due to the increasing 
contribution of the red giant, which is the dominating source in infrared bands. 
In the middle panel  (20170222), the light curves of MWC~560 in B- and V-bands are plotted 
together with the calculated B-V colour.
The right panel are UBV data obtained on 20191025 when the flickering is missing. 
If it exists at all its amplitude in UBV  is $< 0.02$~mag.

\begin{table*}
\centering
\caption{Journal of observations. In the table are given the date 
of observations, the telescope, the filter, UT start and end of the run, detection/non-detection of flickering.    
}  
\label{tab.j}
\begin{tabular}{lllrrlrlllllll}
 \hline
 Date        &    telescope      & bands  &  UT start - UT-end &  detection & \\
 yyyymmdd    &                   &        &  \\
 \hline 

 20091114    &  2.0 Roz          &  U              &  23:59 -- 01:27 &  yes  \\
             &  50/70 Sch        &  B              &  00:21 -- 01:32 &  yes  \\
	     &  60 Bel           &  V              &  00:22 -- 01:27 &  yes  \\
	     &  60 Roz           &  R              &  00:14 -- 01:34 &  yes  \\
 20100111    &  2.0 Roz          &  U, V           &  23:31 -- 01:04 &  yes  \\  
             &  50/70 Sch        &  B              &  23:28 -- 01:10 &  yes  \\ 
	     &	60 Roz           &  R, I           &  23:13 -- 01:01 &  yes  \\ 
 20100315    &  60 Roz           &  V, I           &  18:00 -- 19:15 &  yes  \\ 
 20100317    &  60 Roz           &  V, I           &  18:08 -- 19:14 &  yes  \\ 
 20101229    &  2.0 Roz          &  U              &  22:55 -- 01:28 &  yes  \\
             &  60 Bel           &  V, R           &  22:10 -- 01:30 &  yes  \\
	     &	60 Roz           &  B, I           &  21:53 -- 01:31 &  yes  \\ 
 20110210    &  2.0 Roz          &  U              &  19:18 -- 22:43 &  yes  \\ 
             &  60 Roz           &  B, I           &  19:07 -- 22:45 &  yes  \\ 
             &  60 Bel           &  V, R           &  19:20 -- 22:41 &  yes  \\ 	     
 20110211    &  60 Roz           &  U, B           &  19:35 -- 22:59 &  yes  \\    
             &  60 Bel           &  V, R, I        &  19:34 -- 23:01 &  yes  \\  
 20110212    &  60 Roz           &  U, B, V, R, I  &  20:25 -- 22:47 &  yes  \\   
 20120321    &  60 Bel           &  B, V, R        &  18:27 -- 20:20 &  yes  \\   
 20120323    &  60 Bel           &  B, V, R, I     &  18:06 -- 20:15 &  yes  \\    
 20130303    &  60 Bel           &  B, V, R, I     &  19:35 -- 21:10 &  yes  \\ 
 20130305    &  60 Bel           &  B, V, R, I     &  18:37 -- 20:31 &  yes  \\ 
 20131129    &  60 Roz           &  B, V, R, I     &  00:26 -- 03:31 &  yes  \\ 
 20151118    &  60 Bel           &  B, V, R, I     &  02:10 -- 04:19 &  yes  \\
 20160402    &  60 Bel           &  B, V           &  18:12 -- 19:25 &  yes  \\
 20160405    &  50/70 Sch        &  B              &  17:40 -- 19:11 &  yes  \\ 
 20170222    &  60 Bel           &  B, V           &  22:37 -- 00:09 &  yes  \\
 20180124    &  50/70 Sch        &  B, V, R, I     &  22:18 -- 22:50 &  yes  \\ 
 20191022    &  50/70 Sch        &  B, V           &  00:17 -- 03:49 &  no   \\ 
 20191025    &  50/70 Sch        &  U, B, V        &  00:59 -- 03:19 &  no   \\
 20200201    &  50/70 Sch        &  B              &  19:50 -- 21:15 &  no  \\
 \hline  							      
 \end{tabular}  						      
\end{table*}							      

%
\begin{table*}
\centering
\caption{Data used to calculate the B-V colour of MWC~560.  In the Table are given the date, 
$N_{pts}$ (the number of data points over which B-V colour is calculated), 
average, minimum and  maximum magnitudes in B and V bands.  } 
\label{tab.BV}
\begin{tabular}{lrrrrrrrrlll}
 \hline
date             & $N_{pts}$ &	   mean(B) &	  min(B)  &     max(B)  &    mean(V)   &      min(V)  &    max(V) &   \\
YYYYMMDD         &           &     [mag]   &       [mag]  &     [mag]   &    [mag]     &      [mag]   &    [mag]  &   \\
 \hline   
20091114  &    39  &	11.8557  &   11.799   &  11.921  &  11.2766  &    11.240  &	  11.329  &  \\
20100111  &    52  &	11.5749  &   11.410   &  11.715  &  11.0412  &    10.894  &	  11.163  &  \\
20101229  &   360  &	10.1520  &   10.040   &  10.245  &   9.6801  &     9.587  &	   9.778  &  \\
20110210  &   477  &	10.3856  &   10.224   &  10.580  &   9.9110  &     9.747  &	  10.108  &  \\
20110211  &   186  &	10.4426  &   10.331   &  10.568  &   9.9949  &     9.899  &	  10.116  &  \\
20110212  &   143  &	10.3205  &   10.205   &  10.460  &   9.8805  &     9.776  &	  10.018  &  \\
20120321  &    70  &	10.6389  &   10.496   &  10.752  &  10.1055  &    9.9440  &	  10.205  &  \\
20120323  &    93  &	10.9392  &   10.776   &  11.079  &  10.4181  &    10.227  &	  10.577  &  \\
20130303  &   100  &	10.8227  &   10.755   &  10.888  &  10.3261  &    10.247  &	  10.394  &  \\
20130305  &    79  &	11.0507  &   10.880   &  11.191  &  10.5484  &    10.385  &	  10.662  &  \\
20131129  &    75  &	10.9839  &   10.764   &  11.134  &  10.5579  &    10.331  &	  10.696  &  \\
20151118  &    87  &	10.1774  &   10.032   &  10.420  &   9.6246  &     9.481  &	   9.813  &  \\
20160402  &    88  &	 9.3861  &    9.202   &   9.535  &   8.7897  &     8.605  &	   8.953  &  \\
20170222  &   110  &	10.1601  &   10.055   &  10.276  &   9.6060  &     9.508  &	   9.725  &  \\
20180124  &    27  &	10.2431  &   10.134   &  10.299  &   9.8144  &     9.716  &	   9.868  &  \\
20191022  &   207  &	 9.2703  &    9.254   &   9.293  &   8.9272  &     8.909  &	   8.950  &  \\
20191025  &   114  &	 9.2885  &    9.273   &   9.311  &   8.9387  &     8.917  &	   8.967  &  \\ 
\hline  
 \end{tabular}  
\end{table*}

 \begin{figure*}    
   \vspace{7.0cm}     
   \includegraphics{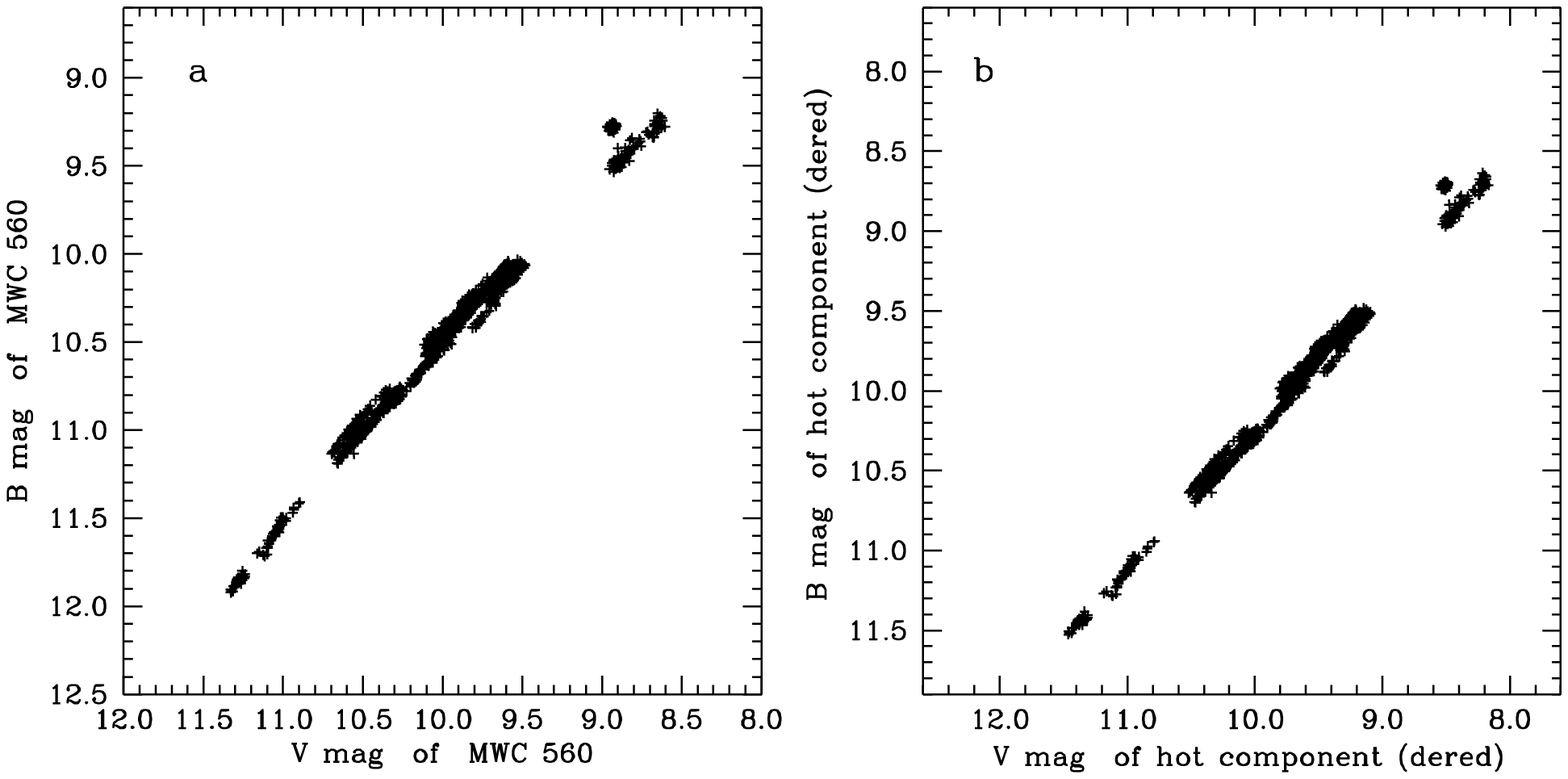} 
   \caption[]{B versus V band magnitude:
      \hskip 0.1cm
      a) observed, 
      \hskip 0.1cm  
      b) calculated for the hot component. } 
   \label{fig.mm} 
   \vspace{7.0cm}      
   \includegraphics{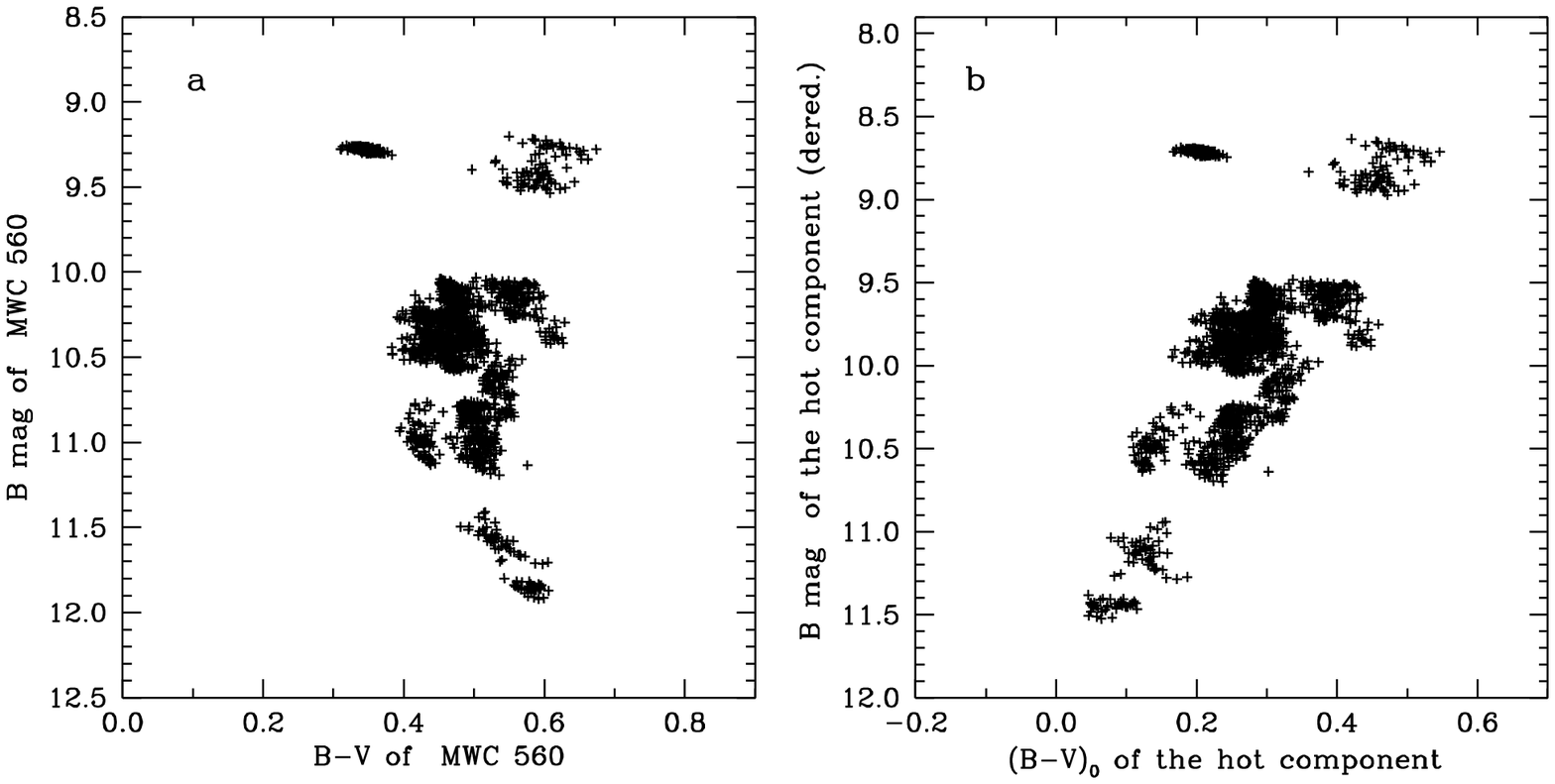}  
   \caption[]{Colour magnitude diagram:  
     \hskip 0.1cm 
      a) observed, 
     \hskip 0.1cm 
      b) calculated for the hot component.  } 
   \vspace{7.0cm} 
   \label{fig.cm2}        
   \includegraphics{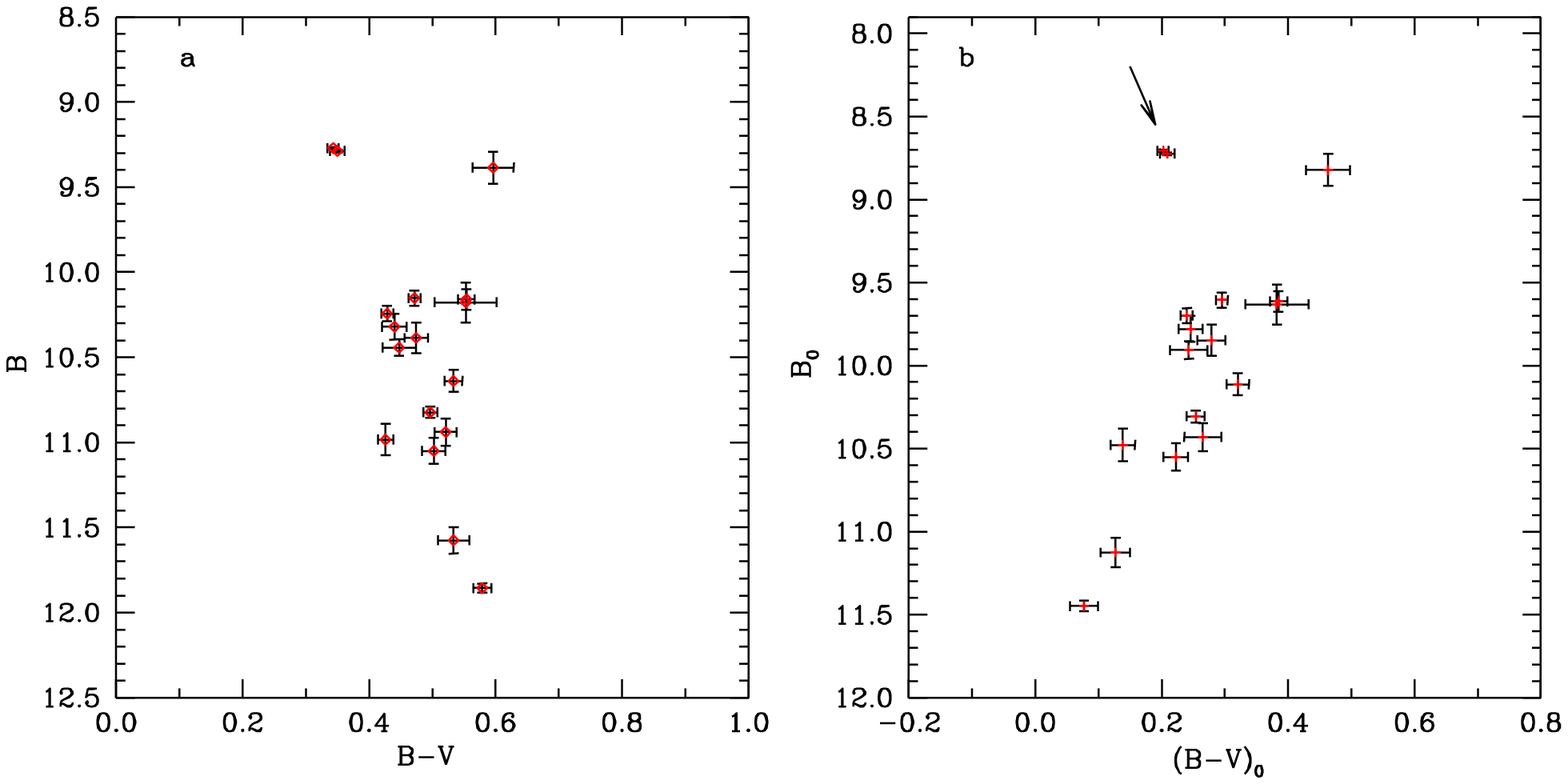}      
   \caption[]{Mean colour magnitude diagram, each point represent one run: 
      \hskip 0.1cm 
         a) observed,
      \hskip 0.1cm  
         b) calculated for the hot component, the arrow indicates the two nights without 
     detectable flickering. }
   \label{fig.cm3}      
 \end{figure*}	     

We have 17 nights with simultaneous observations in B and V bands 
during the period July 2008  - October 2019. 
The $B-V$ colour is calculated for 2307 points in total.
During our observations the brightness of MWC~560 was: \\ 
$9.20 \le  B    \le  11.92 $, \\
$8.60 \le  V    \le  11.33 $,  \\
$0.31 \le  B-V  \le   0.67 $,\\
with  mean B = 10.28,  mean V = 9.81,  mean B-V = 0.47. 
When the flickering exists, its
peak-to-peak amplitude in B band is in the range  0.13 - 0.39 mag. 
 
The journal of observations is given in Table~\ref{tab.j}. 
In Table~\ref{tab.BV} are given the number of data points over which B-V colour is calculated, 
average, minimum and  maximum magnitude in B- and V-bands.

\section{Parameters of the system}
\label{sec.par}

GAIA DR2 (Gaia Collaboration et al. 2018) gives for MWC560  parallax $0.3534 \pm  0.1659$, 
which corresponds to a distance d=2830 pc. 
Schmid et al. (2001) derived distance $2.5 \pm 0.7$ kpc, which agrees with the GAIA value.

Schmid et al. (2001) estimated interstellar extinction 
E(B-V)=0.15 mag  from 2200 \AA\  feature and for the mass donor spectral type  M5.5 III. 
Slightly different values are given earlier by Zhekov et al. (1996)  -- M4.5 III and  E(B-V)=0.23 mag. 
Most likely the extinction is in  the range $0.1 \le E(B-V) \le 0.2 $
in the light of the NaD absorption and dust maps (Lucy et al. 2020). 
Houdashelt, Wyse \& Gilmore (2001) give for  M5.5 III giant colours  
B-V=1.55 and  V-I=2.7. 


For the red giant, Zhekov et al. (1996) estimated  $m_B \sim 14$~mag, which is in agreement with the  
long term light curve of MWC~560 (Doroshenko, Goranskij \& Efimov 1993). With the above colour and extinction, this 
give for the red giant  $m_V \sim 12.30$~mag.
Using J and K-band data from 2MASS All Sky Catalog (J=6.452, K=5.069) 
and 
the extinction law from Savage \& Mathis (1979), 
we obtain A$_J$ = 0.13 and A$_K$ = 0.06. Koornneef (1983) gives an intristic colours 
V-K = 6.7 and V-J = 5.43 for an M5.5~III star. 
Using these parameters, we derive m$_V$ = 12.23 and
m$_V$ = 12.19 using J and K-band respectively.


Hereafter, we assume for the red giant component of MWC~560, $m_V \approx 12.25$ and $m_B \approx 13.94$. These magnitudes 
are used in Sect.~\ref{sec.BV} to estimate the colours of the hot component. 

%
%

\section{Variability in B and V bands}
\label{sec.BV}

In Fig.\ref{fig.mm} we plot B versus V band magnitude.
In the left  panel (Fig.~\ref{fig.mm}a) are the observed magnitudes of MWC~560. 
In the right panel (Fig.~\ref{fig.mm}b) are the de-reddened magnitudes of
the hot component (e.g. the red giant contribution is subtracted using the magnitudes given
in Sect. \ref{sec.par}).  

In Fig.\ref{fig.cm2} we plot colour-magnitude diagrams. 
The colour-magnitude diagram is quite different from that of the cataclysmic 
variable AE Aqr (see Fig.3 in  Zamanov et al. 2017) but it is similar to that 
of the recurrent nova RS~Oph (see Fig.~3,4,5 in Zamanov et al. 2018).  
In the case of AE Aqr all the data occupy a well defined strip. 
Here such a strip is not visible, although the observations from each night 
are placed on a specific position on the diagram.
This indicates  that the flickering behaviour  
of MWC~560 has 
a similar mechanism to that of RS Oph, but different than 
that of AE~Aqr. 

In Fig.\ref{fig.cm3} we plot the calculated mean values for each night (one night -- one point).
The error bars correspond to the standard deviation of the run. 
The left panel is the  observed, the right panel is the hot component (red giant contribution subtracted).
On Fig.\ref{fig.cm3}a, it can be seen  
that the colour of the system is in the range $0.31 \le  B-V  \le   0.67 $,
without clear tendency to become redder or bluer when the brightness changes.

However, there is a  correlation between the mean colour and magnitude of the hot component.
When we use the 15 points, when the flickering exists we calculate
Pearson correlation coefficient 0.91, Spearman's (rho) rank correlation  0.88,
the statistical significance  $p\mbox{--}value = 5.9 \times 10^{-5}$. 
This indicates that the hot component  becomes redder as it gets brighter. 

When we use all 17 points including the two nights without flickering
this correlation weakens: 
Pearson correlation coefficient  0.58, Spearman's (rho) rank correlation 0.45  
$p\mbox{--}value = 0.07$. 
This indicates that the missing flickering is connected with violation
of  colour - brightness relation of the accretion disc 
and/or 
probably changes in its structure/geometry.

\section{Flickering light source}
\label{s.fli}

Bruch (1992) proposed that the light curve of a white dwarf with flickering  can be separated into two parts -- constant light
and variable (flickering) source. 
Following his recipe, 
we calculate the flux of the flickering light source 
as $F_{\rm fl1}=F_{\rm av}-F_{\rm min}$, where $F_{\rm av}$ is the average flux 
during the run and $F_{\rm min}$ is the minimum flux during the run
(corrected for the typical error of the observations).
An expansion of the method is proposed by
Nelson et al. (2011). 
They suggest to use the $F_{\rm fl2}=F_{\rm max}-F_{\rm min}$, where $F_{\rm max}$ 
is the maximum flux during the run. 
In fact, the method of Bruch (1992) evaluates the average brightness of the flickering source, 
while that of Nelson et al. (2011) --  its maximal brightness. 
$F_{\rm fl1}$ and $F_{\rm fl2}$  have been calculated for each band, using the values 
given in Table~\ref{tab.BV} and 
the calibration for a zero magnitude star $ F_0 (B) =6.293 \times 10^{-9}$  erg cm$^{-2}$ s$^{-1}$ \AA$^{-1}$,    $\lambda_{eff}(B)=4378.12$~\AA, 
$F_0 (V) = 3.575 \times 10^{-9}$  erg cm$^{-2}$ s$^{-1}$ \AA$^{-1}$ and  $\lambda_{eff}(V)=5466.11$~\AA\ as given in Spanish virtual observatory 
Filter Profile Service  (Rodrigo et al. 2018, see also Bessell 1979).  

It is worth noting that while the calculated colours of the hot component depend on
the assumed red giant brightness, the parameters of the flickering source are independent on the 
red giant parameters. 

Using method of Bruch (1992), 
we find that in B band the flickering light source  contributes about 
12\%  of the average flux of the system,  with    $ 0.06  \le  F_{\rm fl1} / F_{\rm av} \le 0.20$.
In V band its average contribution is 11\%,  with    $ 0.05  \le  F_{\rm fl1} / F_{\rm av} \le 0.17$. 

Using method of Nelson et al. (2011), 
we find that in B band the flickering light source  contributes about 
22\%  of the maximal flux of the system,  with   $0.11   \le  F_{\rm fl2} / F_{\rm max} \le 0.30 $.
In V band its is about 21\%,  with  $  0.08  \le  F_{\rm fl2} / F_{\rm max} \le 0.29$.

From the amplitude - flux relation (rms-flux relation)  e.g. Scaringi et al. (2015),
we expect that the luminosity of the flickering source will increase as the brightness increases. 
However, it is not a priori clear which parameter --  temperature or radius (or both),  increases.

In Table~\ref{tab.fl} are given the dereddened colour of the flickering source $(B-V)_{01}$ and  $(B-V)_{02}$,  
T$_1$  and  T$_2$ - temperature of the flickering source, 
R$_{1}$  and  R$_2$ - radius the flickering source.

In the calculations we assume that the flickering source is continuum dominated. 
In principle, it is possible that the flickering involves both continuum and lines.  
However, the search for rapid spectral variability related to the flickering 
in MWC~560
shows that there 
are no changes higher than a few per cent level in the optical lines in spite of 0.35 mag flickering 
in B band (Tomov et al. 1995). 
Our simultaneous 5-colour photometry shows  
that the flickering source is well approximated with black body   
in the UBVRI bands (Zamanov et al. 2011b). 
The observations of the near-UV flickering, show that 
near-UV spectral morphology remained constant so the 
near-UV flickering must have originate in a variable continuum (Lucy et al. 2020). 
Bearing in mind the above, 
as well the  discussion in Sect. 6.2 of Sokoloski et al. (2001) about
the difficulty of producing rapid variability from the nebular emission, we
adopt that the flickering of MWC~560 is continuum dominated and  reflects the physical origin
of the variations in the accretion disc around the white dwarf.

\subsection{B-V colour and temperature of the flickering source}
\label{BV.T}

The calculated de-reddened  colours of the flickering light source are given in Table~\ref{tab.fl},  
where $(B-V)_{01}$ is calculated using $F_{av}$ and $F_{min}$, while  $(B-V)_{02}$ is calculated using $F_{max}$
and  $F_{min}$. Typical error is $\pm 0.05$ mag. 

\begin{figure}    
   \vspace{7.8cm}     
   \includegraphics{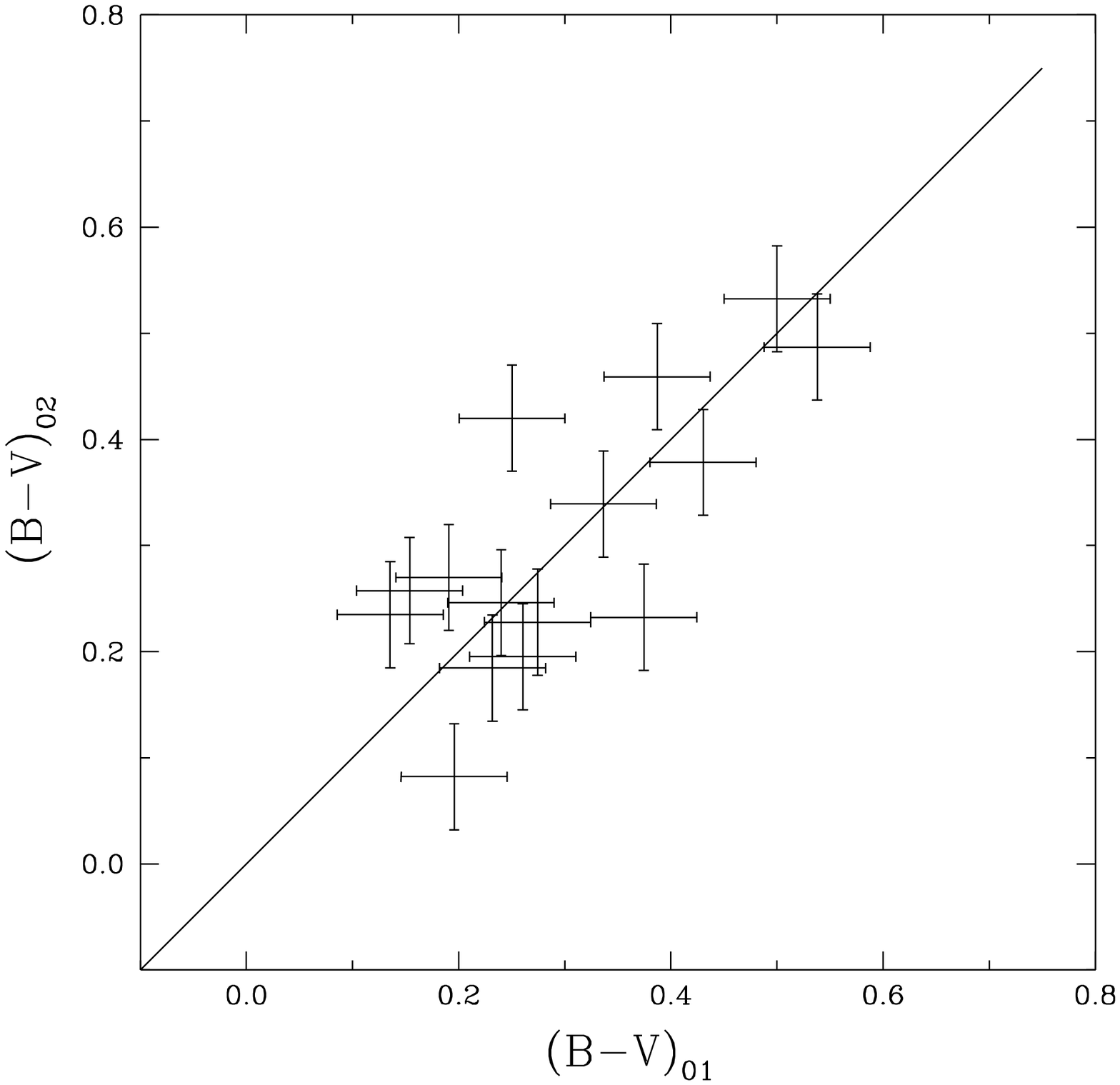}      
   \caption[]{$(B-V)_{02}$ versus $(B-V)_{01}$ --
    there is no systematic shift between the two methods. See Sect.~\ref{BV.T} for details.}
   \label{fig.41}      
   \vspace{7.9cm}     
   \includegraphics{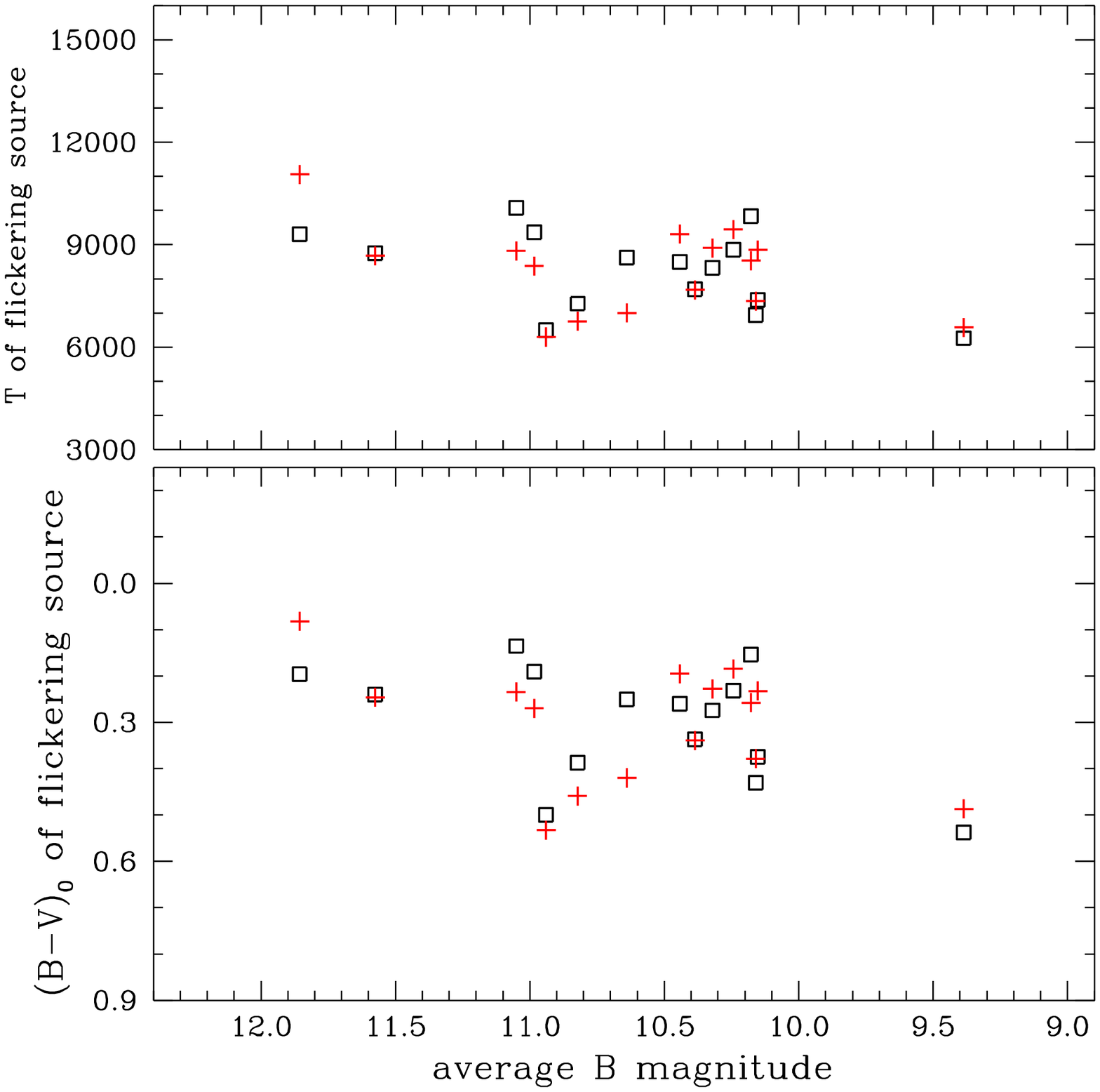}      
   \caption[]{ Temperature and $(B-V)_0$ colour of the flickering source versus the average B magnitude. 
    The squares represent  values calculated by method of 
    Bruch (1992), the plus signs --  derived following Nelson et al. (2011).  }
    \label{fig.42}      
 \end{figure}

In Fig.\ref{fig.41}  we plot $(B-V)_{02}$  versus $(B-V)_{01}$.  
The solid line represents $(B-V)_{02} = (B-V)_{01}$. 
To  check for a systematic shift between the two methods we performed linear least-squares 
approximation in one-dimension $(y = a + b x)$, when both x and y data have errors.
We obtain $a= 0.01 \pm 0.02$ and $ b = 0.95 \pm 0.07$.  
A Kolmogorov-Smirnov test gives Kolmogorov-Smirnov statistic 0.07  and  significance level 0.99.
It means that both methods give similar results and there is not a systematic shift.
The average difference between them is $\approx  0.07$ mag,
which is comparable with the accuracy of our estimations. 
In Fig.\ref{fig.42}, we plot $(B-V)_0$ versus the average B band magnitude. 
We do not detect a statistically significant correlation
between the colour of the flickering source and the brightness of the system.


We calculate the temperature of the flickering source using its dereddened colours 
and the colours of the black body (Straizys, Sudzius \& Kuriliene 1976).
$T_1$ is calculated using $(B-V)_{01}$, and $T_2$ is calculated using  $(B-V)_{02}$. 
The two methods give similar results for
the  temperature of the flickering source as well as for $(B-V)_{0}$. 
The average  values are $T_1 = 8244 \pm 1180$ and  $T_2 = 8241 \pm 1290$.

\subsection{Radius and luminosity of the flickering source}
\label{R.fl}

\begin{figure*}    
   \vspace{8.9cm}     
   \includegraphics{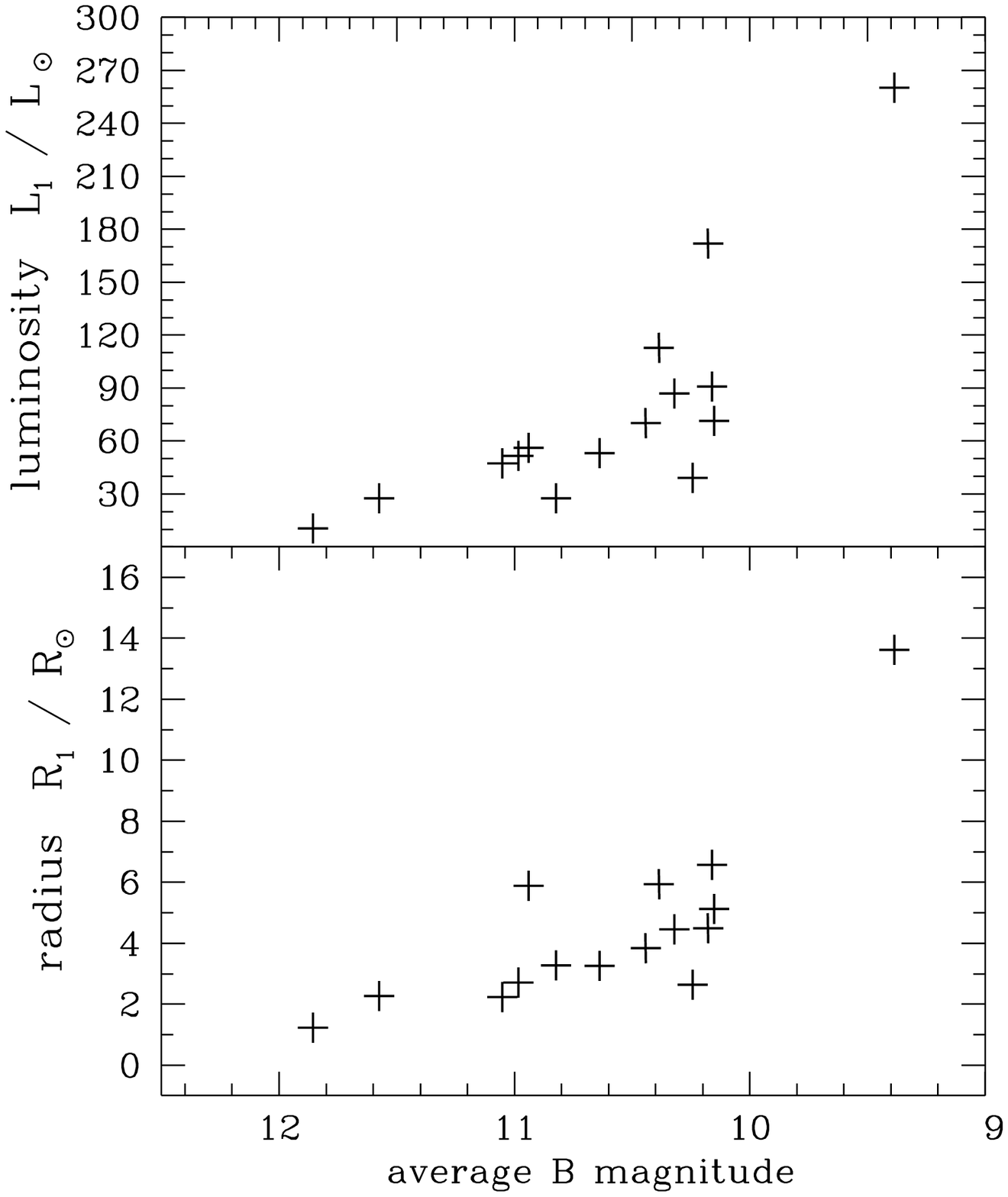}      
   \includegraphics{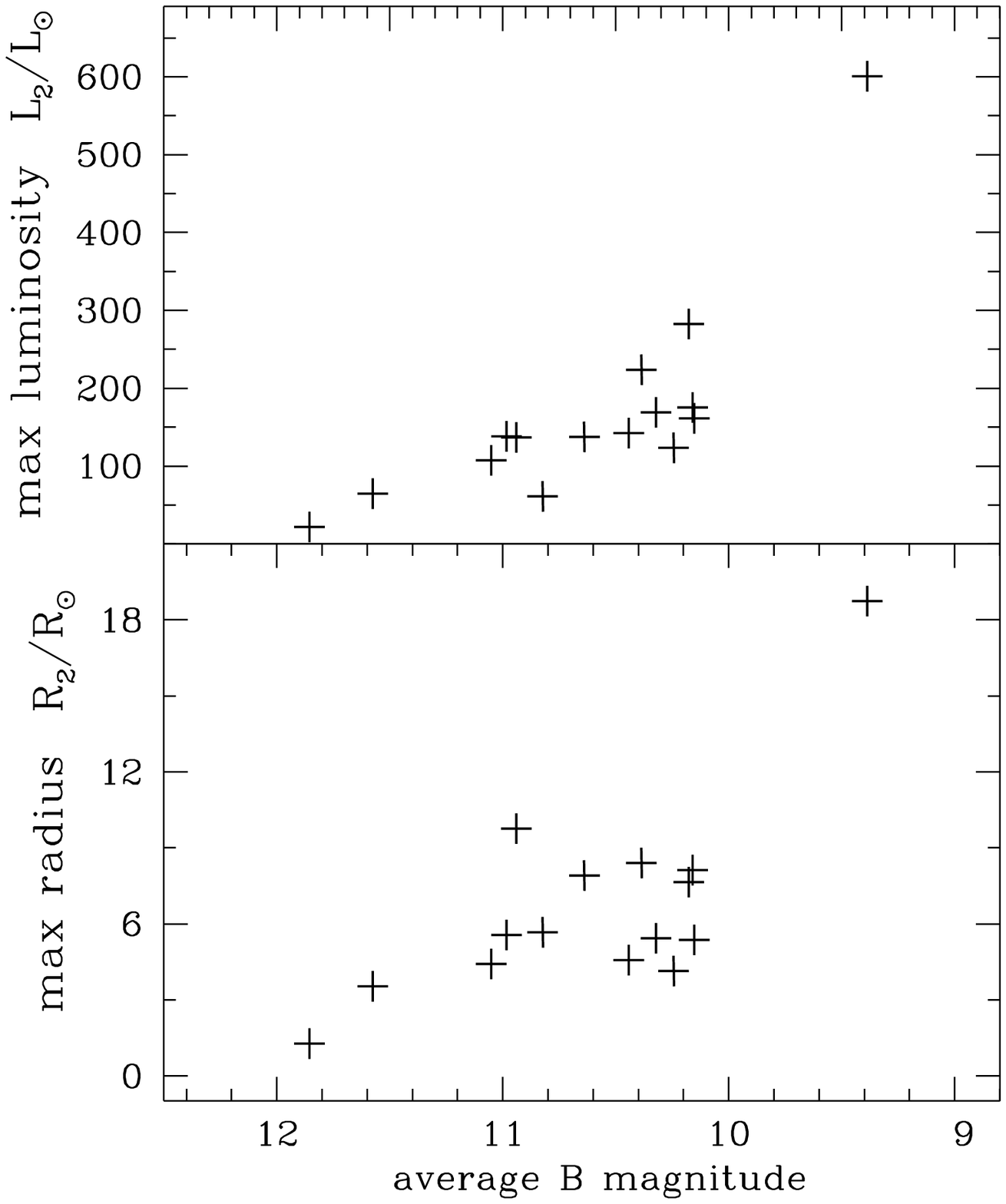} 
   \caption[]{Radius and luminosity of the flickering source of MWC~560 versus the average B band magnitude. 
    The left panels are  average  $L_1$ and $R_1$, the right are the maximum $L_2$ and $R_2$
    (see Sect.~\ref{s.fli}).  }
   \label{fig.LR}      
 \end{figure*}

The radius of the flickering source  is calculated 
using the derived temperature (Sect.~\ref{BV.T}), the
B-band magnitude and assuming  that it is spherically symmetric. 
We obtain  $ 1.2 <  R_1 < 13.6 $~\rsun\
and  $ 1.3 <  R_2 < 18.7 $~\rsun. 
In Fig.~\ref{fig.LR} we plot $R_1$ and  $R_2$ versus the average B band magnitude. It is seen that 
radius of the flickering source increases  when the brightness of the system increases. 

The luminosity of the flickering source  is calculated 
using the derived temperature and the radius given in Table~\ref{tab.fl}: 
$L_{1} = 4 \pi R_{1}^2 \sigma T_1^4$, 
where  $\sigma$ is the Stefan-Boltzmann constant. 
We obtain  10.4~\lsun ~ $<  L_1 < 260$~\lsun,
and 22~\lsun ~ $<  L_2 < 600$~\lsun.

We found a strong correlation  between  the average B band magnitude and $R_1$
with Pearson correlation coefficient 0.765,   Spearman's (rho) rank correlation 0.761,
significance  $p\mbox{--}value = 9.9 \times 10^{-4}$.
The correlation between  B and $R_2$ is weaker. 
There are also strong correlations between  B and  $L_1$ --
Pearson 0.785, Spearman 0.782, $p\mbox{--}value =   5.7 \times 10^{-4}$,
as well as between B and  $L_2$ -- 
Pearson 0.79, Spearman's  0.80,  $p\mbox{--}value =  3.0 \times 10^{-4}$.

\begin{table*} 
\small
\caption{The calculated parameters of the flickering source of MWC~560. 
$(B-V)_{01}$, T$_1$ and  R$_{1}$  are dereddened colour, temperature and radius of the  flickering source calculated following Bruch (1992), 
$(B-V)_{02}$, T$_2$ and  R$_{2}$ --  following Nelson et al. (2011), see Sect.~\ref{s.fli} for details. }
\begin{center}
\begin{tabular}{ccrrc | crrrrrrrrrrrr}
\hline
date       & $(B-V)_{01}$  &  T$_1$   & R$_{1}$  &  &  & $(B-V)_{02}$  & T$_2$   & R$_2$    & \\
YYYYMMDD   &	           &  [K]     & [\rsun]  &  &  &	       & [K]	 & [\rsun]  & \\
\hline
           &               &          &          &  &  &	       &	 &	    & \\   
20091114   &  0.1958       &    9302  &   1.24   &  &  &  0.0821       &   11053 &   1.28   & \\   
20100111   &  0.2400       &    8750  &   2.27   &  &  &  0.2461       &    8674 &   3.55   & \\   
20101229   &  0.3744       &    7380  &   5.14   &  &  &  0.2323       &    8846 &   5.38   & \\  
20110210   &  0.3365       &    7696  &   5.94   &  &  &  0.3391       &    7674 &   8.41   & \\  
20110211   &  0.2604       &    8495  &   3.85   &  &  &  0.1952       &    9310 &   4.57   & \\   
20110212   &  0.2743       &    8321  &   4.46   &  &  &  0.2277       &    8904 &   5.43   & \\   
20120321   &  0.2503       &    8621  &   3.25   &  &  &  0.4199       &    7001 &   7.92   & \\  
20120323   &  0.5001       &    6499  &   5.88   &  &  &  0.5325       &    6297 &   9.76   & \\  
20130303   &  0.3870       &    7275  &   3.29   &  &  &  0.4591       &    6756 &   5.67   & \\   
20130305   &  0.1355       &   10082  &   2.24   &  &  &  0.2348       &    8815 &   4.42   & \\  
20131129   &  0.1907       &    9366  &   2.71   &  &  &  0.2699       &    8376 &   5.55   & \\   
20151118   &  0.1537       &    9829  &   4.50   &  &  &  0.2575       &    8531 &   7.65   & \\   
20160402   &  0.5379       &    6263  &  13.63   &  &  &  0.4870       &    6581 &  18.74   & \\  
20170222   &  0.4304       &    6935  &   6.57   &  &  &  0.3784       &    7347 &   8.13   & \\   
20180124   &  0.2319       &    8851  &   2.64   &  &  &  0.1845       &    9444 &   4.13   & \\  
\\
\hline
\end{tabular}																      
\end{center}																               
\label{tab.fl}
\end{table*}

\section{Discussion}

During the last decades, MWC~560 underwent three optical brightenings - 1990, 
2010-2011 and 2016. 
Multiwavelength observations (including optical, ultraviolet, 
X-ray and radio data) in the last years as well as a model
accounting for the similarities and the differences between 
the brightening events is presented by Lucy et al. (2020). 
The observations of the intranight variability during the 1990 outburst are 
presented and analysed in Tomov et al. (1996)  and Zamanov et al. (2011a).
The present data set covers the period 2009 -- 2019.
The long term light curve (Doroshenko et al. 1993; Leibowitz  \&  Formiggini 2015) indicates that 
during our observations MWC~560 was on average 3-4 times brighter in the optical bands  than before 1989.  

Random fluctuations of the brightness are
observed throughout diverse classes of objects 
that accrete material onto a compact object (white dwarf, neutron star or black hole)  
 -- cataclysmic variables, X-ray binaries,  Active Galactic Nuclei.  
Photoelectric observations identified the flickering as a common characteristic of the accreting white dwarfs
in cataclysmic variables 
(e.g.  Mumford 1966, Henize 1949, Robinson 1973). 
The flickering appears as  stochastic light variations on time-scales of about 10 minutes 
with amplitude  from a few $\times 0.01$ mag to more than one magnitude.
Three different places are considered as source of flickering from accreting white dwarfs --
the accretion disc itself, its outer edge (bright spot), 
and its inner edge (boundary layer).

\subsection{ Bright spot }

The temperature and the size of the bright spot 
are derived for a few 
cataclysmic variables. 
For OY Car, Wood et al (1989) calculated
temperature in the range 8600 -- 15000~K; 
Zhang \& Robinson (1987) for U Gem - $T = 11600 \pm 500$ K;
Robinson, Nather \& Patterson (1978) give $T = 16000$~K 
for the bright spot in WZ Sge.  
For IP~Peg three estimates exist:   
Marsh (1988) -- $T = 12000 \pm 1000$~K,  Ribeiro et al. (2007) -- 6000-10000 K, 
Copperwheat et al. (2010) -- 7000 - 13000 K.
The temperature of the optical flickering
source of MWC~560 is in the range $6200 < T_{fl} < 10100$~K (see Sect.4.1), which is similar to
the temperature of the bright spot of cataclysmic variable stars. 

The bright spot is produced by the impact of the stream on the outer parts of the
accretion disc. In case of Roche-lobe overflow this stream is coming from the inner Lagrangian point $L_1$.
If the red giant in MWC~560 does not fill its Roche lobe, the white dwarf accretes material from its wind. 
In this case accretion cone and accretion wake will be formed, e.g. Fig.~4 of  Ahmad, Chapman \& Kondo (1983). 
The stream formed in the accretion wake should be similar to that formed from Roche-lobe overflow.  
The luminosity of the bright spot is approximately (Shu 1976;  Elsworth \& James 1982): 
\begin{equation}
 L_{bs} \approx  \frac{1}{2} \;  V_\perp ^2 \; \dot M_{acc},  
\label{eq.hs}
\end{equation}
where $\dot M_{acc}$ is the mass accretion rate and $V_\perp$ is the inward component 
of the stream's velocity at the impact with the disc. 
Eq.~\ref{eq.hs}  indicates that when the mass accretion rate
increases, the luminosity of the spot also must increase.
In addition, our results (Sect.~\ref{R.fl}) indicate that when the mass accretion
rate increases the radius of the bright spot (if it is the source
of flickering) also increases, while its temperature remains almost constant.

\subsection{Temperature in the accretion disc}
\label{Tdisc}

The broad-band variability is often attributed to 
{\bf (i)} inward propagating fluctuations driven by
stochastic variability in the angular momentum transport mechanism (Lyubarskii 1997);
{\bf (ii)}  turbulence, vortexes in the disc (e.g. Dobrotka et al. 2010; Kurbatov \& Bisikalo 2017);
{\bf (iii)} spiral structures in the disc (e.g. Baptista \& Bortoletto 2008). 
The timescales of changes of the overall structure of the accretion disc 
are longer compared to the local fluctuating processes in the flow that 
can generate the flickering. In this way 
the dynamical time scale variability of the flickering light source do not
change the overall structure of the accretion disc.
Considering the entire disc structure, 
the temperature in the disc can be approximated with 
the radial temperature profile of a steady-state accretion disc (e.g.  Frank, King \& Raine 2002):
\begin{equation}
T_{eff}^4=\frac{3 G \dot M_{acc} M_{wd}}{8 \pi \sigma R^3} 
\left[ 1-\left(\frac{R_{wd}}{R}\right)^{1/2} \right] ,
\end{equation}
$R$ is the radial distance from the white dwarf.
We assume   $ M_{wd} = 0.9$~\msun\ and $R_{wd}= 6\times 10^8$~cm (Zamanov, Gomboc, \& Latev 2011; Lucy et al. 2020)
and  mass accretion rate of about  $5 \times  10^{-7}$ ~\msun~yr$^{-1}$ (Schmid et al. 2001).

Using the parameters for MWC~560, 
a temperature  $6300 \le T_{fl} \le 10000$~K 
(the temperature of the flickering light source as given in  
Table~\ref{tab.fl}) should be achieved at a distance 
$R\approx 1.2-2.5$ \rsun\ from the white dwarf.
If the accretion disc itself 
is the place for the origin of the flickering of MWC~560, then it 
comes at distance $R \sim 2$~\rsun\ from the white dwarf.

\subsection{Boundary layer}
 
The boundary layer between the white dwarf and the inner edge of the accretion disc should have 
temperature $\ge 10^5$~K (e.g. Mukai 2017). The derived temperature of the flickering source 
is considerably lower than that expected from the boundary layer. 
If the boundary layer is optically thick,
in this case the radius of the flickering source measured in the optical bands (Fig.~\ref{fig.LR})
could represent the radius up to which the emission generated from the boundary layer 
is reprocessed by the inner parts of the accretion disc and  the accretion disc corona.

\subsection{Disappearance of the flickering}

The first indication that flickering of symbiotic stars disappears sometimes 
was found for the recurrent nova T~CrB (Bianchini \& Middleditch 1976).
In CH Cyg the flickering was missing for  more than 3 years  (Stoyanov et al. 2018).
In RS~Oph the flickering disappeared after the nova outburst and re-appeared 241 days later 
(Worters et al. 2007). 
In MWC~560 the flickering was visible in all
observations obtained between 1984 and May 2018 (Tomov et al. 1996, Zamanov et al. 2011a,b; Lucy et al. 2020). 
It disappeared in October 2018 (Goranskij et al. 2018) and is not visible 
in our  observations obtained in October 2019 and February 2020. 
What can be the reason for disappearance of the flickering : 
\begin{enumerate}
 \item If the source of the flickering is bright spot than it means  
that for some reason  
$V_\perp \approx 0$  (see Eq.~\ref{eq.hs}). In other words 
at the impact point,  the stream has velocity approximately equal to the velocity of the outer disc edge. 

 \item If the source of the flickering is the accretion disc, the disappearance 
means that the stochastic fluctuations disappear and the disc becomes stable (non-fluctuating).  

 \item Goranskij et al. (2018) proposed that
a common envelope is formed due to the transit of the system to a dynamical mode of accretion with an increased rate. 
The accretion matter filling the Roche lobe of the compact companion blocked the jets and 
overlapped the direct visibility of the companion, so flickering was deleted.
\end{enumerate}
In MWC~560 at the disappearance of the flickering the hot component becomes bluer, 
but its brightness in B band remain high (see Fig.~\ref{fig.cm3}b, where the arrow indicates the observations
without flickering).  This could be an indication that the accretion disc 
becomes smaller and/or hotter. 

In X-rays MWC~560 is  $\beta/ \delta $ type, i.e.
with two X-ray thermal components -- soft and  hard (Luna et al. 2013).   
The  soft emission is most likely produced in a colliding-wind region, 
and the hard emission is most likely
produced in boundary layer between accretion disc and white dwarf. 
The changes in the X-ray emission can give us clues why the flickering  disappeared.

\section{Conclusions}

We  report quasi-simultaneous observations 
of the flickering variability of the jet-ejecting symbiotic star MWC~560
in  17 nights during the period  November 2011 - October 2019.  
The colour-magnitude diagram, B versus B-V, shows 
that when the flickering exists, 
the hot component of the system becomes redder as it gets brighter.

For the flickering source we find that it has colour  in the range  $0.14 < B-V < 0.40$,    
temperature in the range  $6300 - 11000$~K, 
and radius  in the range $1.2 - 18$~\rsun. 
The estimated temperature is similar to that of the bright spot of cataclysmic variables.
We do not find a correlation between the temperature of the 
flickering and the brightness.  
However, we do find  strong correlations 
(1) between B band magnitude and the average radius of the flickering source --
as the brightness of the system increases  the size of the flickering source also increases;
(2) between B band magnitude and the luminosity  of the flickering source --
as the brightness of the system increases  the luminosity of the flickering source also increases.  
When the flickering disappeared in 2019, the B-V colour of the hot component becomes bluer
and its brightness in UBV remains high. 

The behaviour of the hot component and flickering source in MWC~560  
should provide useful input for theoretical modeling of accretion in symbiotic type binaries.

\vskip 0.3cm 

{\bf  Acknowledgements:  We dedicate this paper to the memory of Prof Toma Tomov (1953 - 2019)}, who initiated the 
observations of MWC~560 at Rozhen Observatory.
This work was supported by the grant   KP-6-H28/2 "Binary stars with compact object" (Bulgarian National Science Fund).


\end{document}